\documentclass[twocolumn,prl,amsmath,amssymb]{revtex4}
\usepackage{graphicx}
\usepackage{epstopdf}
\usepackage{amsmath}
\usepackage{threeparttable}
\usepackage{multirow}
\usepackage{color}
\begin{document}
\title{Tetrahedral amorphous carbon resistive memories with graphene-based electrodes}
\author{A. K. Ott, C. Dou, U. Sassi, I. Goykhman, D. Yoon, J. Wu, A. Lombardo, A. C. Ferrari}
\affiliation{Cambridge Graphene Centre, University of Cambridge, Cambridge CB3 0FA, UK}
\begin{abstract}
Resistive-switching memories are alternative to Si-based ones, which face scaling and high power consumption issues. Tetrahedral amorphous carbon (ta-C) shows reversible, non-volatile resistive switching. Here we report polarity independent ta-C resistive memory devices with graphene-based electrodes. Our devices show ON/OFF resistance ratios$\sim$4x$10^5$, ten times higher than with metal electrodes, with no increase in switching power, and low power density$\sim$14$\mu$W/$\mu$m$^2$. We attribute this to a suppressed tunneling current due to the low density of states of graphene near the Dirac point, consistent with the current-voltage characteristics derived from a quantum point contact model. Our devices also have multiple resistive states. This allows storing more than one bit per cell. This can be exploited in a range of signal processing/computing-type operations, such as implementing logic, providing synaptic and neuron-like mimics, and performing analogue signal processing in non-von-Neumann architectures.
\end{abstract}
\maketitle
\section{\label{In}Introduction}
There are two main types semiconductor random access memories (RAM): static (SRAM)\cite{memory-book} and dynamic (DRAM)\cite{memory-book}. Both are volatile\cite{memory-book}, i.e. they lose the stored information when power is off\cite{memory-book}. SRAMs are currently used in central processing unit (CPU) registers due to their fast write/erase times$\sim$0.3ns\cite{Jeong2012}. However, typical SRAM cells comprise six transistors\cite{electronics-book,memory-book} and require larger footprint\cite{memory-book}, e.g. a cell size$\sim$0.04$\mu$m$^2$ for a 6 transistors (6T) SRAM\cite{SRAM-footprint}, compared to DRAM cells, with a footprint$\sim$0.024$\mu$m$^2$\cite{IEEE-white-paper}. Due to this, and the scaling limit of the gate oxide in transistors, further decreasing the SRAMs footprint is difficult\cite{memory-book,Jeong2012}. DRAMs store information in an integrated capacitor\cite{memory-book}. They require periodic refreshing to regenerate the data, due to their gradual discharge due to leakage currents\cite{Jeong2012}, resulting in increased power consumption, one third of which due to refreshing\cite{DRAM-refresh}. Flash memories are non-volatile\cite{memory-book}, and rely on electrical charge stored by a floating gate\cite{memory-book}, making scaling below 16nm difficult\cite{IEEE-white-paper}, because the oxide layer, which acts as tunnel barrier\cite{memory-book} surrounding the floating gate, cannot be made arbitrarily thin, otherwise the charge stored in the floating gate will be lost\cite{memory-book}. This also limits the write and erase data rates to$\sim$1ms/0.1ms for NAND-Flash drives\cite{Jeong2012}.

Resistive RAMs (RRAM) are non-volatile memories based on a change in resistance initiated by electric fields\cite{Fu2011,Fu2014,Jeong2012}. They are promising candidates for next generation non-volatile memory technology, due to the high speed$\sim$ns\cite{Kreupl2008,Koelmans2016}, low operational power$\sim$pJ\cite{Lanza2017, Kreupl2008}, and integrability with conventional complementary metal-oxide-semiconductor, CMOS, processes\cite{Waser2007,Waser2009,Sawa2008,Wong2012,Jo2009,Lanza2017}. The key performance indicators of RRAMs are\cite{Lanza2017,memory-book}: operation voltages; power consumption; endurance, i.e. the number of times non-volatile memory (NVM) can be switched on/off before one of the states becomes irreversible; ON/OFF ratio R$_{\text{ON}}$/R$_{\text{OFF}}$ and data retention, i.e. the time before a state change without application of voltage happens.

Resistive switching (RS) in metal-oxide RRAMs based on the formation and annihilation of localized conductive filaments (CFs) has been extensively studied\cite{Sawa2008,Wong2012,Lanza2017}. One of the most critical challenges is the electro-forming process to activate RS\cite{Jeong2012, Waser2009, Sawa2008,Wong2012}. A forming voltage is needed to introduce enough oxygen deficiencies to create the CF for the first time\cite{Sawa2008}. The SET and RESET processes, i.e. changes from non-conducting to conducting state and vice versa, are attributed to rupturing and re-forming of CF sections close to either top or bottom reactive metal electrodes, due to oxygen migration and electrode reduction/oxidation\cite{Sawa2008}. However, only a portion of the oxygen deficiencies near the electrodes can be recovered during RESET, through an interfacial redox effect in the subsequent switching cycles\cite{Ielmini2016,Waser2009, Sawa2008,Wong2012}. As a result, the forming voltage is usually much larger than the operational voltage, as it is necessary to form the biggest CF part, and R$_{\text{ON}}$/R$_{\text{OFF}}$ is degraded, so that ON and OFF states become indistinguishable, since the device's resistance cannot be switched back to the initial resistive state\cite{Ielmini2016}. These effects do not only lead to an increased complexity for the memory periphery circuitry due to the large forming voltage\cite{Ielmini2016,Waser2009, Sawa2008,Wong2012}, but could also limit the integration in devices with large storage capacity\cite{Wong2012}. Therefore, research is ongoing to devise ``forming-free'' RRAMs, with R$_{\text{ON}}$/R$_{\text{OFF}}$ of at least 10$^3$\cite{IEEE-white-paper} to keep leakage currents to a minimum\cite{IEEE-white-paper}.

RRAMs are usually integrated in circuits as crossbar arrays, i.e. the top electrode is rotated by 90 degrees with respect to the bottom one\cite{Linn2010}. This layout offers scaling potential only limited by the minimum lithographic feature size, and the ability to generate multilevel stacking memories\cite{Sawa2008}. Crossbar arrays consist of a lower and an upper plane of parallel wires, running at right angles\cite{Burr2014}. If both wires and spaces between them have a width F, defined as the minimum lithographic feature size\cite{IEEE-white-paper}, then the area per connection is a=4F$^2$\cite{Burr2014}, where a=cell size/half pitch$^2$\cite{IEEE-white-paper}. By eliminating the access transistor integrated in the RRAM crossbar it is possible to achieve the smallest theoretical cell size 4F$^2$\cite{Shevgoor2015,Zhang2015}. However, a small cell size without access transistor requires an R$_{\text{ON}}$/R$_{\text{OFF}}$ of at least 10$^3$ to guarantee a stable and reliable performance\cite{Shevgoor2015,Zhang2015}, which cannot be achieved with all materials and technologies\cite{Lanza2017}. RRAMs are also promising to bridge the access time gap between fast ($\sim$60ns\cite{IBM-gap}), but volatile, DRAMs and non-volatile, but slow ($\sim$20$\mu$s-5ms\cite{IBM-gap}) storage devices.

RS in amorphous carbons has been extensively studied\cite{Fu2011,Fu2014,Chai2010, Dellmann2013, Kreupl2011,Kreupl2008,Sohn2014,Xu2014,Sebastian2011, Koelmans2016, Wright2017, Lanza2017}. Carbon-based RS is particularly attractive for the following reasons. First, it offers high-temperature stability without degradation, with data retention demonstrated at 300$^{\circ}$C for 600mins\cite{Fu2011}, making it suitable for applications in harsh environments, such as in the automotive\cite{Kreupl2014} and aerospace sectors\cite{Kreupl2014}. Second, compared to metal oxides and other RS materials, carbon-based RRAMs are down-scalable, with cell sizes in the sub-100nm range demonstrated\cite{Koelmans2016}, with fast ($<$10ns\cite{Kreupl2008, Koelmans2016}) and low power ($\sim$pJ range\cite{Kreupl2008,Koelmans2016}) switching. Third, R$_{\text{ON}}$/R$_{\text{OFF}}$ and threshold switching voltage (V$_{th}$) can be tuned and optimized for different applications by controlling the ratio between sp$^3$ and sp$^2$ carbons or by the inclusion of hetero-atoms, such as oxygen\cite{Koelmans2016,Santini2015}, hydrogen\cite{Dellmann2013, Sebastian2011} and nitrogen (see e.g. Refs.\cite{Wright2017,Chen2014}), even though this makes them not monoatomic. When the sp$^3$ fraction is higher than 50\%, amorphous carbons are called tetrahedral-amorphous carbon, ta-C\cite{Casiraghi2007c, Ferrari2004hd, Fallon1993,Polo2000,Ferrari2000}.

Several advantages of introducing layers of graphene into metal-oxide RRAMs have been reported\cite{Qian2014,Tian2013,Yang2014}, such as suppressed tunneling current\cite{Qian2014}, monitoring oxygen movement\cite{Tian2013}, built-in selector effect acting as a switch for the RRAM cell\cite{Yang2014}, and reduced power consumption\cite{Qian2014}. Refs.\cite{Qian2014,Tian2013, Yang2014} explored the possibility of introducing chemical vapor deposition (CVD) grown graphene, whereas Refs.\cite{Qian2014,Tian2013} used single layer graphene (SLG), and Ref.\cite{Yang2014} used 3 layers of graphene in oxide based RRAMs, with a 47 times reduced power consumption\cite{Tian2013}. In oxide based RRAMs, oxygen ions migrate under an electric field from the oxide to the top electrode (anode) leading to the formation of conductive oxygen vacancy filaments during the SET process, while for the RESET oxygen ions move from the anode back to the oxide layer as result of the application of a field with opposite voltage polarity. Refs.\cite{Qian2014,Tian2013,Yang2014} suppressed the tunneling current through TiO$_x$\cite{Tian2013} by using SLG as bottom electrode. Ref.\cite{Tian2013} monitored the oxygen movement at the interface between top contact and HfO$_x$ by inserting SLG, which served as oxygen barrier, preventing oxygen ions to move into the electrode material\cite{Tian2013}. Ref.\cite{Yang2014} reported that multilayer graphene, MLG, can be used as a selector or switch, analogous to threshold or insulator-to-metal-transition (IMT) switches. They transferred 3 layers on top of each other, as top and bottom electrodes in a MLG/Ta$_2$O$_{5-x}$/TaO$_y$/MLG structure. The MLG bottom electrode was oxidized as it was heated to 400$^{\circ}$C and subjected to an oxidative Ar/O$_2$ plasma\cite{Yang2014}, and served as a selector element or switch for a RRAM cell\cite{Yang2014}. Thus graphene could be used to optimize performance of other RS systems.

Here we introduce SLG in ta-C based metal-insulator-metal (MIM) devices, with voltage polarity-independent RS, with SET and RESET achieved either by unipolar or bipolar switching, and forming-free RS, with R$_{\text{ON}}$/R$_{\text{OFF}}\sim$10$^5$. By using electrodes based on CVD grown SLG to cover the whole chip size of up to 2$\times$2cm$^2$, we find that R$_{\text{ON}}$/R$_{\text{OFF}}$ increases by one order of magnitude without increasing V$_{th}$ and current for the SET process. We investigate the underlying mechanism by comparing a control device without SLG, and devices with fixed size of top metal electrodes and varying SLG size, as well as fixed SLG size and varying top electrode size. We use a quantum point contact (QPC) model to describe RS. The increase in R$_{\text{ON}}$/R$_{\text{OFF}}$ is attributed to suppressed tunneling when using SLG-based electrodes, due to the low SLG density of states (DOS) near the Dirac point. Our data show that SLG is an excellent candidate to tackle the biggest disadvantage of crossbar arrays, i.e. leakage currents, and paves the way for replacing conventional RRAM architectures enabling scaling down to 4F$^2$ cell sizes\cite{Shevgoor2015,Zhang2015}. We also detect multiple resistive states, making our devices attractive for multilevel data storage (MDS), with more than one bit per cell, and increased storage density.
\section{\label{Exp} Experimental}
\begin{figure*}
\centerline{\includegraphics[width=150mm]{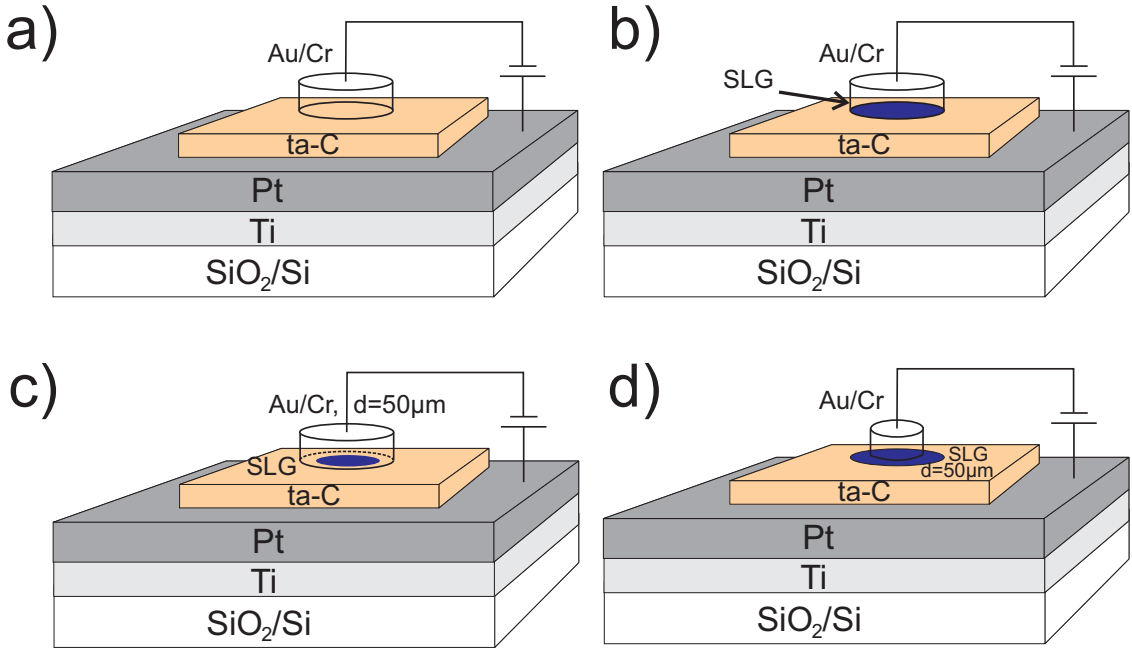}}
\caption{Control device (a) without and (b) with SLG. (c) Au/SLG/ta-C/Pt cells with different SLG size from 0 to 40$\mu$m in steps of 10$\mu$m and fixed top Au electrode diameter of 50$\mu$m. (d) Au/SLG/ta-C/Pt cells with fixed SLG size, with diameter 50$\mu$m, and different diameters Au electrodes, between 20 to 40$\mu$m, in steps of 10$\mu$m.}
\label{fig:dev-structures}
\end{figure*}
\begin{figure}
\centerline{\includegraphics[width=80mm]{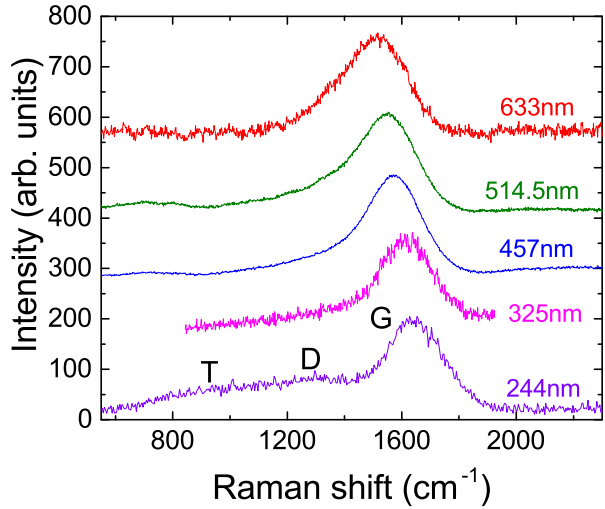}}
\caption{Multiwavelength Raman spectra recorded on a 15nm ta-C film.}
\label{fig:Fig1}
\end{figure}
We consider 3 types of devices, Fig.\ref{fig:dev-structures}. The first is used as reference and has a bottom electrode (Ti/Pt)/ta-C/top electrode (Cr/Au) structure either without, Fig.\ref{fig:dev-structures}a), or with SLG inserted between ta-C and top electrode, where SLG has the same size as the top electrode, Fig.\ref{fig:dev-structures}b). The second and third have either larger top contact compared to SLG, Fig.\ref{fig:dev-structures}c), or smaller, Fig.\ref{fig:dev-structures}d). In each of them a 10nm Ti adhesion layer and a 40nm Pt layer are sputtered on a SiO$_2$(285nm)/Si substrate. A 15nm ta-C film is then deposited through a stainless steel shadow mask using a single-bend filtered cathodic vacuum arc\cite{Fallon1993}. This thickness is chosen to avoid delamination from the Pt substrate.

The ta-C film is characterized by multiwavelength Raman spectroscopy at 244, 325, 457, 514.5 and 633nm using a Renishaw InVia spectrometer equipped with a Leica DM LM microscope. All carbons show common features in their Raman spectra in the 800 to 2000cm$^{-1}$ region, the so called G and D peaks, which lie at$\sim$1560 and 1360cm$^{-1}$, respectively, for visible excitation, and the T peak at$\sim$1060cm$^{-1}$, which can be detected for UV excitation\cite{Ferrari2004,Ferrari2002}. The G peak is due to the bond stretching of all pairs of sp$^2$ atoms in both rings and chains\cite{Ferrari2000}. The D peak is due to the breathing modes of sp$^2$ atoms in rings\cite{TuinstraKoenig,Ferrari2000,Thomsen2000}. The T peak is assigned to the C-C sp$^3$ vibrations\cite{Ferrari2001, Piscanec2005}. An empirical three-stage model was developed to describe the Raman spectra of carbon films measured at any excitation energy\cite{Ferrari2000,Ferrari2001-aC,Ferrari2004}. The evolution of the spectra is understood by considering an amorphisation trajectory, starting from graphite. The main factor affecting peaks position, width and intensity is the clustering of the sp$^2$ phase. This can in principle vary independently from the sp$^3$ content, so that for a given sp$^3$ content and excitation energy, we can have a number of different Raman spectra\cite{Ferrari2000,Ferrari2004} or, equivalently, similar Raman spectra for different sp$^3$ contents\cite{Ferrari2000,Ferrari2004}. For UV excitation, an increase in clustering lowers the G peak position, Pos(G). However, in visible Raman the G peak does not depend monotonically on cluster size. If two samples have similar Pos(G) in visible Raman but different ones in UV Raman, the sample with the lower Pos(G) in the UV has higher sp$^2$ clustering\cite{Ferrari2000,Ferrari2004}. A multi-wavelength Raman analysis is thus important to fully characterize the samples. A very useful parameter is then the G peak dispersion Disp(G). This is defined as the slope of the line connecting Pos(G) measured at different excitation wavelengths\cite{Ferrari2004}. Another useful parameter is the Full Width at Half Maximum of the G peak, FWHM(G). Both FWHM(G) and Disp(G) increase as the disorder increases, at every excitation wavelength\cite{Ferrari2000,Ferrari2004}. Thus, Disp(G) allows one to estimate the Young's modulus\cite{Ferrari2004}, density\cite{Ferrari2000d} and sp$^3$ content\cite{Ferrari2004hd, Ferrari2000d,Ferrari2004}. Fig.\ref{fig:Fig1} plots representative spectra measured on the 15nm thick ta-C film on Pt/Ti/SiO$_2$/Si. We have Disp(G)$\sim$0.33cm$^{-1}$/nm, I(T)/I(G)$\sim$0.32, corresponding to sp$^3\sim$60\%.
\begin{figure}
\centerline{\includegraphics[width=0.4\textwidth]{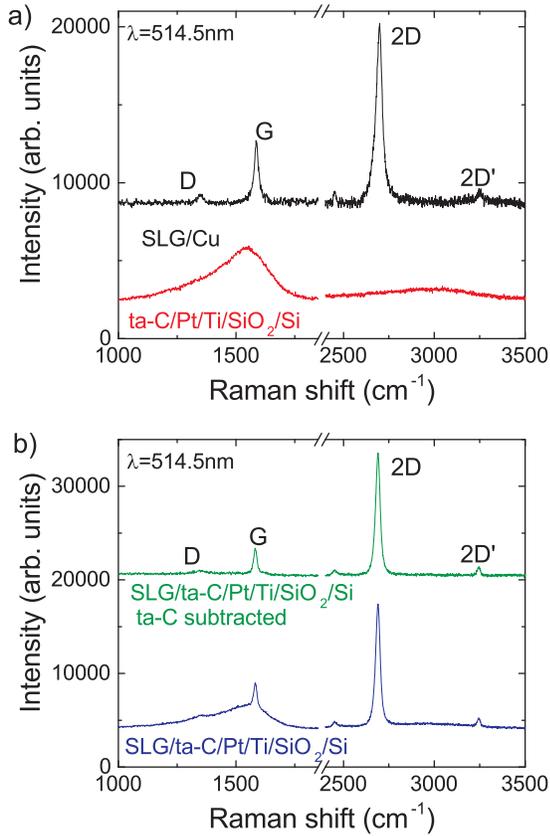}}
\caption{Raman spectra at 514.5nm of a) as-deposited ta-C (red) and as-grown SLG on Cu (black). b) SLG/ta-C stack (blue) and SLG/ta-C after subtraction of the ta-C spectrum (green)}
\label{fig:switching-Raman}
\end{figure}

In a control device, a Au(60nm)/Cr(3nm) electrode is placed on the ta-C film by thermal evaporation and lift-off, Fig.\ref{fig:dev-structures}a. Inert metals, such as Pt and Au, are used as electrodes to ensure that the RS in the ta-C based cell does not result from the diffusion of metal ions and metal filament formation into the ta-C or from interfacial redox effects of the electrodes, unlike RS in oxides\cite{Ielmini2016,Waser2009,Sawa2008,Wong2012}. RS in ta-C is usually assigned to the formation of an sp$^2$ CF in the sp$^3$ matrix between top and bottom electrode, driven by Joule heating\cite{Sebastian2011,Koelmans2016,Bachmann2017}. Au is used to form the metal/SLG contact for SLG, because the SLG DOS is unchanged when Au is placed on top as electrode\cite{Ifuku2013}. This is due to Au's weak influence on SLG's DOS\cite{Ifuku2013}. Ref.\cite{Ifuku2013} reported that the quantum capacitance of a SLG device with Au electrodes arranged in a transfer length method configuration exhibits an ambipolar behavior, suggesting that Au has little influence on the SLG DOS.

SLG is grown by CVD on a 35$\mu$m-Cu foil loaded into a hot wall tube furnace as for Ref.\citenum{Bae2010}. The Cu foil is annealed in a hydrogen atmosphere (H$_2$, 20sccm) at 1000$^{\circ}$C for 30 mins to reduce the Cu oxide component\cite{Bae2010} and to increase the Cu grain size\cite{Bae2010}. Growth is initiated by adding 5sccm CH$_4$ to the H$_2$ flow. After 30 minutes, the substrate is cooled in vacuum (1mTorr) to RT and then unloaded from the reactor. Fig.\ref{fig:switching-Raman}a plots the Raman spectrum of the as-grown material on Cu, after the non-flat PL background of Cu is removed\cite{Lagatsky2013}. The 2D peak can be fitted with a single Lorentzian with FWHM(2D)$\sim$40cm$^{-1}$ indicating SLG\cite{Ferrari2006}. Pos(G) and FWHM(G) are$\sim$1587and$\sim$21cm$^{-1}$, respectively, and the 2D to G intensity and area ratios are I(2D)/I(G)$\sim$3 and A(2D)/A(G)$\sim$5.6, respectively. I(D)/I(G)$\sim$0.14, corresponding to a defect concentration n$_D\sim$3.4x10$^{10}$cm$^{-2}$\cite{Cancado2011,Bruna2014}. SLG is then placed on the ta-C film by a polymethyl methacrylate (PMMA)-based wet transfer\cite{Bonaccorso2012,Bonaccorso2010}. To monitor the SLG quality before and after transfer, the ta-C background signal is measured under the same conditions and subtracted point-by-point by using the Si substrate signal as reference for normalization. Fig.\ref{fig:switching-Raman}b shows that, after transfer, Pos(G) and FWHM(G) are$\sim$1582 and$\sim$16cm$^{-1}$, while I(2D)/I(G) and A(2D)/A(G) are$\sim$4.9 and 8.5, respectively. Pos(2D) and FWHM(2D) are$\sim$2689 and 30cm$^{-1}$. These data point to very small, if any, doping$\ll$100meV\cite{Das2008,Bruna2014,Pisana2007} of SLG on ta-C. I(D)/I(G)$\sim$0.1 indicates that no significant amount of defects has been introduced during transfer.

Our devices are prepared with Au$/$Cr top electrodes diameters of 10, 20, 30, 40 and 50$\mu$m, defined by lithography. O$_2$ plasma reactive-ion etching (RIE) is then used to remove any SLG regions left uncovered. As a result, SLG has the same shape as the electrodes on top of it, Fig.\ref{fig:dev-structures}b. This is one type of reference sample. The second and third set of devices are prepared with two lithography steps to shape the SLG and metal electrodes size. In one case, SLG is shaped by lithography into circular areas with fixed 50$\mu$m diameter, followed by the fabrication of Au(60nm)/Cr(3nm) top electrodes, also having circular shapes, but with diameters of 20, 30, 40 and 50$\mu$m, directly on top of SLG by thermal evaporation and lift-off, Fig.\ref{fig:dev-structures}c. In the other devices, SLG is shaped into different circular areas with varying diameters starting from no SLG to 10, 20, 30, 40 and 50$\mu$m, while the top Au electrode diameter is kept at 50$\mu$m, Fig.\ref{fig:dev-structures}d.

To characterize the devices electrically, I-V curves are recorded on a Keysight CascadeMicrotech probe station using a Keysight B1500A semiconductor analyzer. Ta-C is deposited through a shadow mask, this enables the areas where Pt was protected by the shadow mask to be accessible to the probes.
\section{\label{Res} Results and discussion}
\begin{figure}
\centerline{\includegraphics[width=90mm]{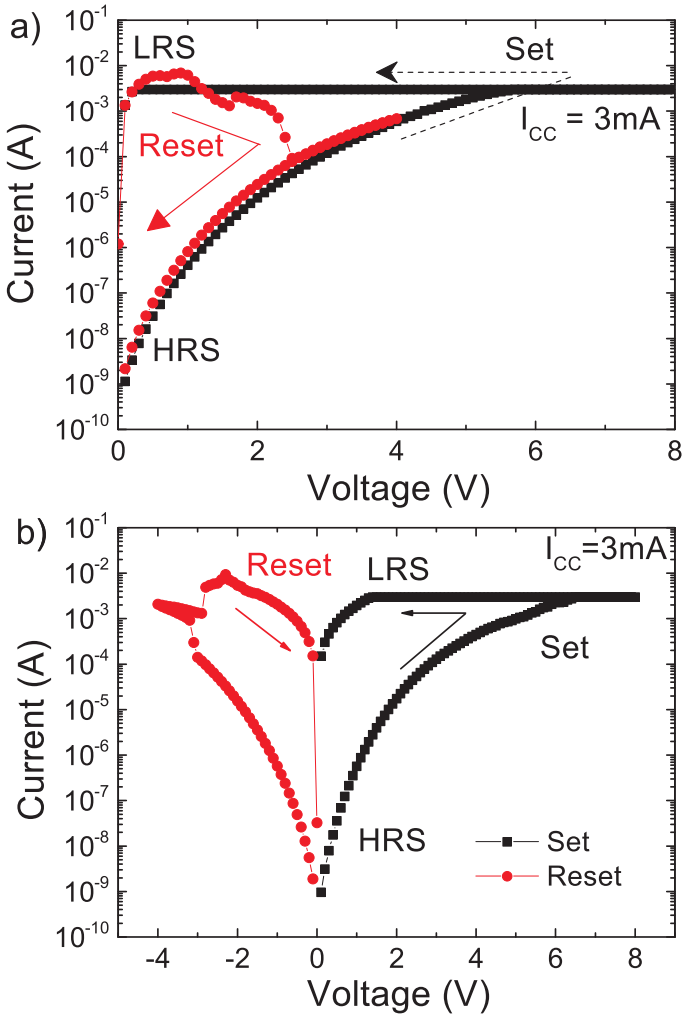}}
\caption{Typical (a) unipolar and (b) bipolar RS I-V curves of the Au/SLG/ta-C/Pt device with circular top electrodes (d=20$\mu$m). The arrows show the directions of voltage sweeping in the (black) SET and (red) RESET process. A compliance current (I$_{CC}$) of 3mA is used in the SET process to protect from hard breakdown. The device under test is that in Fig.\ref{fig:dev-structures}b. The voltage is applied to the Pt bottom electrode.}
\label{fig:Fig2}
\end{figure}
\begin{figure}
\centerline{\includegraphics[width=90mm]{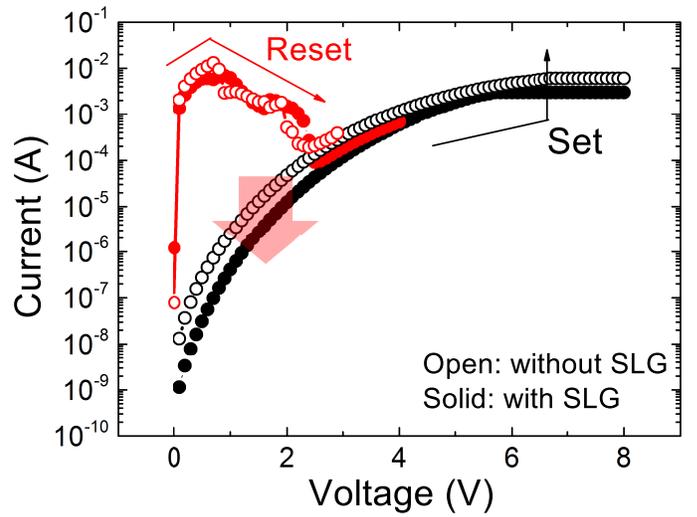}}
\caption{RS curves for (open circles) Au/ta-C/Pt and (solid circles) Au/SLG/ta-C/Pt cells with d=20$\mu$m. 5 and 3mA compliance currents are used for the device without and with SLG. The voltage is applied to the bottom Pt electrode.}
\label{fig:Fig3}
\end{figure}
Fig.\ref{fig:Fig2} plots typical RS I-V curves for an Au/SLG/ta-C/Pt device (d=20$\mu$m) operated in unipolar mode, i.e. with only positive voltage sweep, Fig.\ref{fig:Fig2}a, and bipolar, where opposite voltage polarities are used for SET and RESET, Fig.\ref{fig:Fig2}b. Fig.\ref{fig:Fig2}b shows that RS is independent on the polarities. The device is initially in a high resistance state (HRS) and can be set to a low resistance state (LRS) by sweeping the DC voltage from 0 to 8V, and reset back by a sweeping the voltage from 0 to 4V. A 3mA compliance current, i.e. a fixed maximum limit of current, is also used in the SET process to protect the device from irreversible or hard breakdown, where the CF would be too thick to be ruptured, thus making the RESET process impossible. Fig.\ref{fig:Fig2} also shows that R$_{\text{ON}}$/R$_{\text{OFF}}\sim4\cdot10^5$ at 0.2V, without electro-forming, unlike metal oxides\cite{Ielmini2016,Waser2009, Sawa2008,Wong2012}.

RS in ta-C is different from metal oxides, where it is assigned to electrochemical processes near the interface between oxide and electrode\cite{Ielmini2016,Waser2009, Sawa2008,Wong2012} and depends on polarity\cite{Sawa2008,Wong2012}. Given the single elemental nature of our devices, in our case RS originates from the structural changes inside the ta-C, with no forming process. Forming-free RS was reported in organic materials, such as poly(ethylene glycol dimethacrylate) (pEGDMA)\cite{Lee2015} and assigned to sp$^2$ CFs\cite{Lee2015}. In our devices, while the formation of a sp$^2$ CF during SET can be attributed to electric field induced migration of sp$^2$ carbon sites and clusters in the surrounding sp$^3$ matrix\cite{Fu2011,Fu2014,Chai2010, Xu2014,Sebastian2011} and the transition from sp$^3$ to sp$^2$\cite{Koelmans2016} due to a local current annealing\cite{Koelmans2016, Kreupl2008}, the dissolution of the CF in RESET is most likely caused by a thermal fuse effect in the filament, due to the Joule heat generated by high local current density\cite{Fu2011,Fu2014,Chai2010, Xu2014,Sebastian2011, Koelmans2016,Bachmann2017}, with T$\sim$1600K, as calculated by using atomistic simulations in Ref.\cite{Koelmans2016}. Ref.\citenum{Koelmans2016} pointed out the importance of heat dissipation from top and bottom electrodes. This helps the SET process by inducing the formation of chains of sp$^2$ clusters, which eventually merge to form a single conjugated CF, while heat dissipation in the CF direction, as well as T gradients, are crucial for RESET\cite{Koelmans2016}. RS in ta-C is also different from phase change materials (PCM), where a change from amorphous to crystalline and back is achieved by heating\cite{Raoux2014, Wuttig-PCM}, while local melting and rapid quenching form amorphous regions\cite{Raoux2014, Wuttig-PCM}.
\begin{figure}
\centerline{\includegraphics[width=80mm]{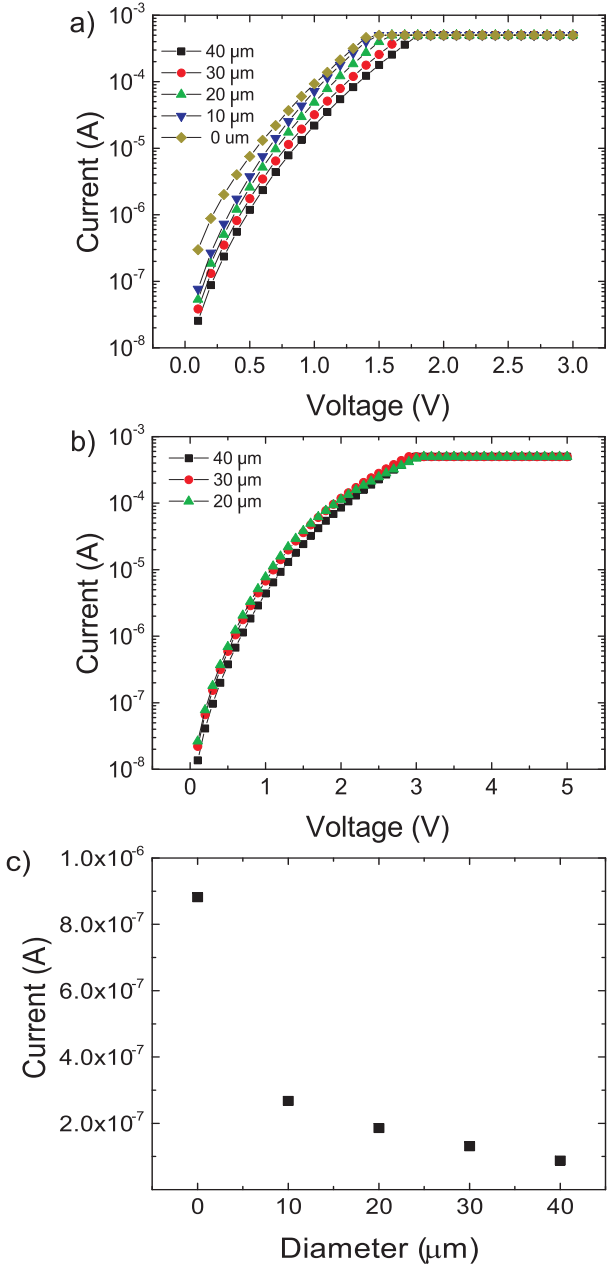}}
\caption{a),b) I-V plots for Au/SLG/ta-C/Pt cells with different d$_{Au}$ and d$_{SLG}$. a) d$_{Au}$=50$\mu$m, d$_{SLG}$ from 0 to 40$\mu$m in steps of 10$\mu$m using the device structure in Fig.\ref{fig:dev-structures}c. b) d$_{SLG}$=50$\mu$m, d$_{Au}$ from 20 to 40$\mu$m in steps of 10$\mu$m using the device structure in Fig.\ref{fig:dev-structures}d. c) Current at 0.2V as function of d$_{SLG}$ for d$_{Au}$=50$\mu$m.}
\label{fig:Fig4}
\end{figure}

In order to investigate the influence of the interfacial SLG, the RS characteristics of the devices with and without SLG are compared in Fig.\ref{fig:Fig3}. Although the two devices show similar RS behavior, that with SLG has lower current by a factor$\sim$4 at the HRS, especially for voltages$<$0.5V. To clarify the reason of the decreased HRS current, Au/SLG/ta-C/Pt cells with different d$_{Au}$ and d$_{SLG}$ are tested. Fig.\ref{fig:Fig4}a plots the I-V characteristics of devices with fixed d$_{Au}$ and different d$_{SLG}$. We keep d$_{Au}$=50$\mu$m, while d$_{SLG}$ ranges from 40 to 0$\mu$m, i.e. no SLG, in steps of 10$\mu$m. In order to ensure that SLG is fully covered by Au and minimize the effect of misalignment between the two steps of lithography, d$_{SLG}$ is kept$<d_{Au}$. Fig.\ref{fig:Fig4}a shows that the HRS current decreases with increasing d$_{SLG}$. Fig.\ref{fig:Fig4}c plots the current at 0.2V for increasing d$_{SLG}$ with fixed d$_{Au}$=50$\mu$m. The HRS current decreases for increasing d$_{SLG}$, reducing  the current by about one order of magnitude for d$_{SLG}$=40$\mu$m, with d$_{SLG}$=10$\mu$m already showing a pronounced effect with a reduction in current of a factor of 4, Fig.\ref{fig:Fig4}c. This leads to an increase in R$_{\text{ON}}$/R$_{\text{OFF}}$ by one order of magnitude from$\sim$10$^5$ to $\sim$10$^6$. Fig.\ref{fig:Fig4}b plots the I-V curves of the devices with d$_{SLG}$=50$\mu$m but different d$_{Au}$. When d$_{Au}$ changes from 40 to 20$\mu$m, the current does not change significantly.

These results indicate that (1) the change of the series resistance due to a different Au/SLG contact area hardly influences the HRS current, and (2) SLG leads to a R$_{\text{ON}}$/R$_{\text{OFF}}$ increase by an order of magnitude. Additionally, the Au electrode has weak interaction with SLG, i.e. it does not change the SLG DOS\cite{Ifuku2013}. We attribute the decreased HRS current of the SLG-based device to suppressed tunneling current from a-C to the contact, due to the low DOS of undoped SLG near the Dirac point. Indeed, Raman measurements on the transferred SLG show that it has very low doping, with E$_F$$\ll$100meV.
\begin{figure*}
\centerline{\includegraphics[width=180mm]{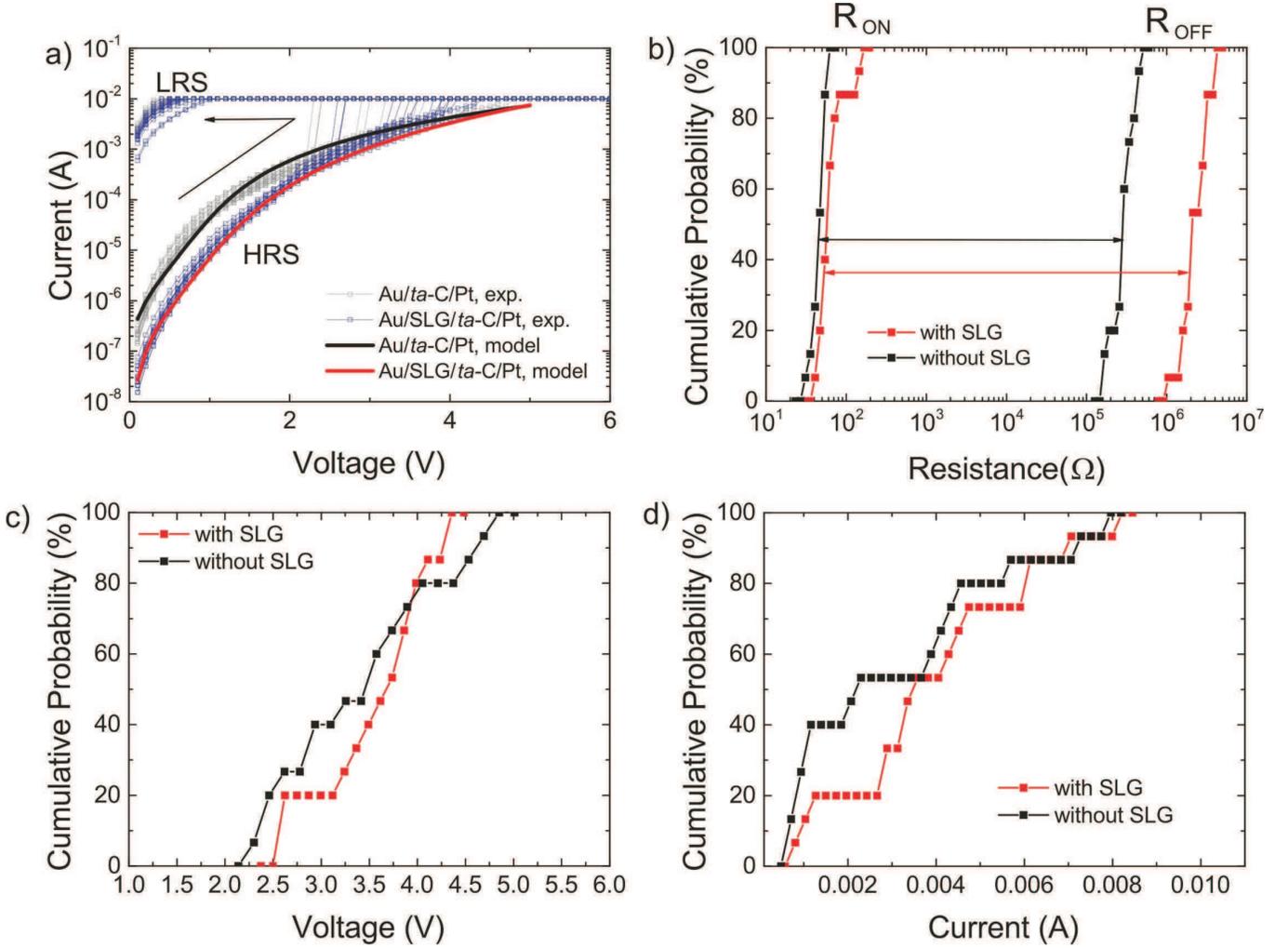}}
\caption{a) (symbols) I-V curves of SET process for devices with SLG-based and metal electrodes, for I$_{CC}$=10mA. (Lines) calculated I-V at HRS from the QPC model. (b) R$_{\text{ON}}$ and R$_{\text{OFF}}$ at 0.2V. (c) Operation voltage. (d) Current required for the SET process. The Y axis in b,c,d reports the integrated probability distribution.}
\label{fig:Fig5}
\end{figure*}
\begin{figure}
\centerline{\includegraphics[width=80mm]{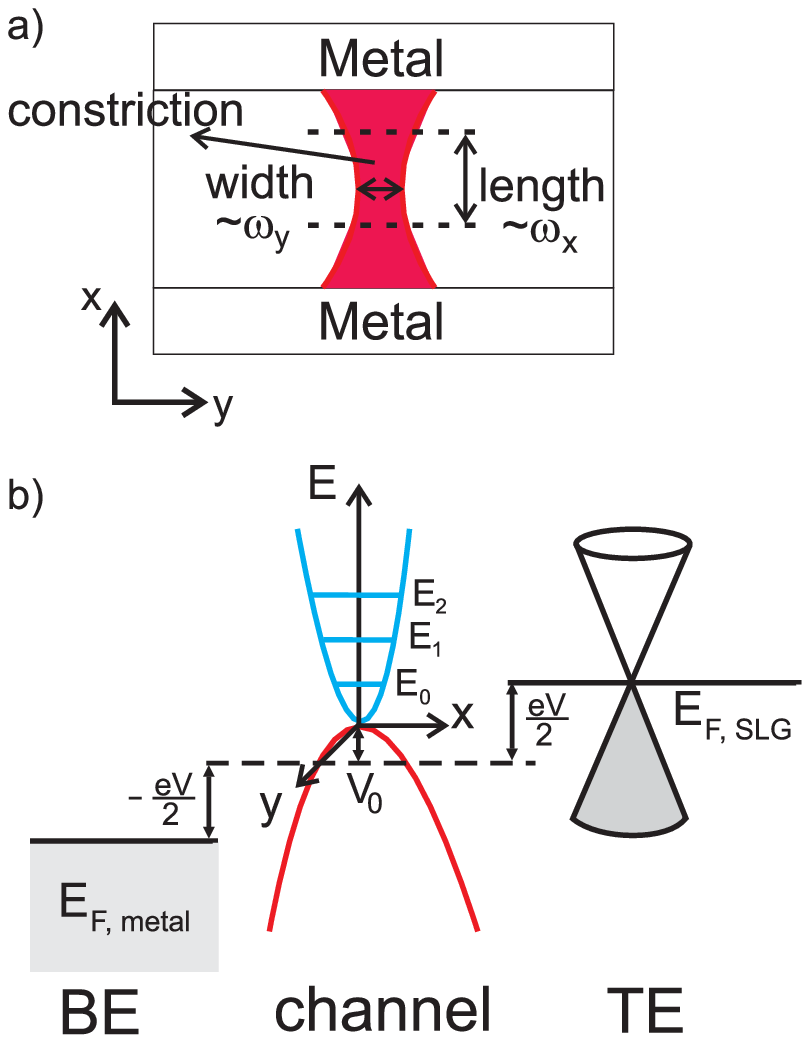}}
\caption{a) Schematic of CF (red) with a spatial constriction. The tunneling current increases with increasing constriction width or decreasing length. Width and length of the constriction are related to the fitting parameters $\omega_y$ and $\omega_x$. b) Band diagram of RS system based on the QPC model with SLG electrode. E$_0$, E$_1$, and E$_2$ are energy levels of the tunneling channels, due to the constriction in the y direction.}
\label{fig:Fig6}
\end{figure}

We analyze 15 different devices with SLG-based and metal electrodes to investigate the SLG influence on a) R$_{\text{ON}}$/R$_{\text{OFF}}$ b), operational voltage, c) current and d) device-to-device variation. The results are summarized in Fig.\ref{fig:Fig5}. This further verifies that SLG suppresses the HRS current at low bias$<$0.5V, with an order of magnitude increase of R$_{\text{ON}}$/R$_{\text{OFF}}$. Having a high R$_{\text{ON}}$/R$_{\text{OFF}}$ is a key requirement for scaling cell sizes down to 4F$^2$\cite{Shevgoor2015,Zhang2015}, as this allows one to remove the access transistor\cite{Shevgoor2015,Zhang2015,Linn2010}. At the same time, the switching voltage and current where the resistance shows an abrupt change in the I-V curve, indicating the change from HRS to LRS, are the same for devices with and without SLG, resulting in a power density$\sim$14$\mu$W/$\mu$m$^2$ for the SET process. This points to the voltage drop on SLG being negligible with no change in operational power density. This can be understood by considering the series connection of the SLG quantum capacitance, i.e. the additional series capacitance related to the DOS\cite{Xia2009}, and the ta-C capacitance. The real part of the dielectric function of ta-C, $\epsilon_1$ can be related to the refractive index as well as an interband effective electron mass, $m^{*}/m=0.87$\cite{Ferrari2000d}, with $n^2(0)=\epsilon_1(0)=(1-m^{*}/m)^{-1}$ and n being the refractive index\cite{Ferrari2000d}. We have $\epsilon_1(0)\sim$7.7 for ta-C using $m^{*}/m=0.87$. Therefore the capacitance of the d=15nm ta-C thin film used here is: C/A=$\epsilon_1(0)\cdot\epsilon_0/d\approx$0.45$\mu$F/cm$^2$. This is at least one order of magnitude smaller than the minimum quantum capacitance of SLG near the Dirac point, which is usually$>2\mu$F/cm$^2$\cite{Xia2009}. From our Raman measurements E$_F\ll$100meV. This corresponds to a charge carrier concentration n$_{ch}=(E_F/\hbar v_F)^2(\pi)^{-1}$, n$_{ch}\ll$0.09cm$^{-2}$. With this, the SLG quantum capacitance can be estimated as C$_Q=2e^2 \sqrt{n_{ch}}/\hbar v_F \sqrt{\pi}<$8.2$\mu$F/cm$^{2}$\cite{Xia2009}, where e=1.6$\cdot$10$^{-19}$C is the elementary charge, n$_{ch}$ is the charge carrier concentration and $v_F$=1.1$\cdot10^6$m/s is the SLG Fermi velocity\cite{Novoselov2005,Zhang2005}. Since these two capacitances are in series, the larger voltage drop is at the smallest one, i.e. 95\% of the voltage drop is at the ta-C layer.

Table \ref{tab:benchmark} compares literature data on R$_{\text{ON}}$/R$_{\text{OFF}}$ and power density for devices based on amorphous carbon (a-C), ta-C, hydrogenated amorphous carbon (a-C:H) and amorphous carbon oxide (a-CO$_x$). This shows that our devices have better R$_{\text{ON}}$/R$_{\text{OFF}}$ compared to RS devices based on SLG and other layered materials\cite{Lanza2017}, while having very low power consumption, proving that SLG is beneficial for optimizing the device performance.
\begin{table}
\caption{Power density and R$_{\text{ON}}$/R$_{\text{OFF}}$ for carbon-based RRAMs.}
\centering
\label{tab:benchmark}
\begin{tabular}{l| c| c }
\multirow{2}{*}{device structure} & power density & R$_{\text{ON}}$/R$_{\text{OFF}}$ \\
&  $\mu$W/$\mu$m$^2$ &\\\hline\hline
W/a-C/Pt ~\cite{Fu2014}         & 1.3x10$^3$ & 1x10$^3$ \\
W/a-C/W ~\cite{Kreupl2008}          & 4.2x10$^2$ & 1x10$^4$ \\
CNT/a-C/CNT ~\cite{Chai2010}      & 8.2x10$^7$ & 1x10$^1$ \\
TiN/a-C:H/Pt ~\cite{Dellmann2013}     & 6.4x10$^5$& 1x10$^2$ \\
TiN/a-C:H/Pt ~\cite{Sebastian2011}     & 6.3x10$^2$ & 1x10$^2$\\
W/a-CO$_x$/Pt ~\cite{Santini2015}       & 2.3x10$^4$ & 1x10$^3$\\
Cu/a-C:N/Pt ~\cite{Chen2014}      & 1.4x10$^0$ & 1x10$^2$\\
W/ta-C/Pt ~\cite{Zhuge2010}        & 3.8x10$^5$ & 1x10$^4$\\
Au/SLG/ta-C/Pt [this work] & 1.4x10$^1$ & 1x10$^5$\\\hline
\end{tabular}
\end{table}

The RS characteristics of ta-C based MIM devices can be analyzed by using the QPC model based on the Landauer theory for mesoscopic conductors\cite{Degraeve2010,Degraeve2012, Miranda2010}. This is based on the idea that the CF behaves as a quantum wire\cite{Degraeve2010,Degraeve2012, Miranda2010}, and was used to describe RS in oxides\cite{QPC-description}. It considers the CF as a 1 dimensional parabolic potential well with an additional parabolic potential barrier along the filament, defining a saddle surface in the x, y, E space given by $E(x,y)=eV_0-0.5 m \omega^2_x x^2 + 0.5 m \omega^2_y y^2$, V$_0$ describes the E$_F$ position in the constriction and $\omega_x$ [Hz] and $\omega_y$ [Hz] are related to and determine the constriction length and width\cite{Degraeve2010,Degraeve2012}.
This model assumes a parabolic potential shape, because it is a good approximation for the saddle-point potential distribution between adjacent conducting sites\cite{Miranda2010, Tekman1991}. The saddle surface with a parabolic potential models the CF constriction in y-direction and the parabolic potential barrier in x-direction\cite{Degraeve2012}. In this model, the current flowing through the CF is limited by a constriction in the middle of the CF as for Fig.\ref{fig:Fig6}a. Thus, the change of the resistive states can be attributed to the widening or narrowing/rupturing of the constriction. The HRS is described by a tunnel barrier height, above the energy of the injected electrons, while for LRS a transmission coefficient T(E)$\approx1$ is assumed, describing the low resistance and high conductance state\cite{Miranda2010}. Fig.\ref{fig:Fig6}b depicts the energy diagram of the RS system under a given applied voltage, V. The constriction leads to a series of tunneling channels with discrete energy levels E$_n$. The total tunneling current, I$_{tot}$, is equal to the sum of the tunneling currents in each channel\cite{Degraeve2010,Degraeve2012}:
\begin{eqnarray}
I_{tot}(V)\sim&    & \nonumber \\
\int^{eV/2}_{-eV/2}&DOS_B\left(E+\frac{eV}{2}\right)DOS_T\left(E-\frac{eV}{2}\right)T(E)dE&\:\:\:\:\:\:
\label{eq:1}
\end{eqnarray}
where $T(E)=\sum^N_{j=0}\left\{1+exp\left[-2\pi\left(E-E_j\right)/\hbar \omega_x\right]\right\}^{-1}$= $\sum^N_{j=0}T_n(E)$, DOS$_B$(E) and DOS$_T$(E) are the DOS of the bottom and top electrodes, T(E) is the total transmission probability of electrons, j is the number of channels, and T$_j$(E) is the transmission probability of the j$^{\text{th}}$ channel. The tunneling current at a given voltage is thus determined by DOS$_B$(E), DOS$_T$(E), and T(E). While DOS$_B$(E) and DOS$_T$(E) are inherent properties of the bottom and top electrodes, T(E) is related to the physical dimensions of the constriction. T(E) decreases with increasing $\omega_x$ or decreasing $\omega_y$\cite{Degraeve2010, Degraeve2012}.

Using Eq.\ref{eq:1} the fitted I-V curves of the HRS for Au/ta-C/Pt and Au/Gr/ta-C/Pt devices are shown by the solid lines in Fig.\ref{fig:Fig5}a. This gives $\omega_x$=1.0$\cdot$10$^{15}$Hz, $\omega_y$=4.4$\cdot$10$^{14}$Hz, V$_0$=0.15eV. Ref.\citenum{Degraeve2010} found similar $\omega_x$=6x10$^{14}$Hz and $\omega_y$=3.8x10$^{14}$Hz with V$_0$=0.25V when simulating a TiN/HfO$_2$/Pt device, showing that the switching processes based on CF formation and rupture are comparable. Several approximations are made to simplify the calculation and illustrate the influence of SLG on the leakage current. The DOS for a single sub-band in a quantum well system can be approximated by$\sim$m$_e^{*}/\pi \hbar^2$\cite{QW-book}, where m$_e^{*}$ is the effective electron mass and $\hbar=h/2\pi$, with h the Planck constant. For Au, m$_e^{*}/$m$_e$=1.1 with m$_e$ the electron mass. This gives DOS$\sim$10$^{18}$eV$^{-1}$m$^{-2}$ for Au. We thus set the Au DOS to a constant$\sim$10$^{18}$eV$^{-1}$m$^{-2}$. For SLG DOS(E)$=2E/\pi\hbar^2v_F^2$\cite{CastroNeto2009}, where $\hbar$ is the reduced Planck constant\cite{Novoselov2005, Zhang2005}. Since the ta-C capacitance of 0.45$\mu$F/cm$^{2}$ is much smaller than SLG, the E$_F$ shift in SLG when a voltage is applied is ignored, as the largest voltage drop is in ta-C. E$_F$ pinning might happen at the interface of SLG with Au\cite{Ando-book}, this means that the SLG E$_F$ shift with applied voltage can be neglected. Furthermore, the same fitting parameters that describe the constriction part of the CF formed in ta-C, i.e. the CF dimensions, are used for both devices with and without SLG ($\omega_x$=1.0$\cdot$10$^{15}$Hz, $\omega_y$=4.4$\cdot$10$^{14}$Hz, and V$_0$=0.15eV), while the SLG influence in the tunneling part is included by taking the SLG DOS into account to model the I-V curves. With this, the modeled I-V curves of devices without and with SLG match the measured I-V curves, Fig.\ref{fig:Fig5}a. This indicates that, due to the low SLG DOS near the Dirac point, SLG suppresses tunneling in the metal/ta-C/metal system, especially$<$0.5V, thus increasing R$_{\text{ON}}$/R$_{\text{OFF}}$. As the voltage increases, the tunneling currents in the devices with and without SLG become closer because, with increasing bias, a higher DOS becomes available in SLG leading to an increase in tunneling or leakage current.
\begin{figure}
\centerline{\includegraphics[width=90mm]{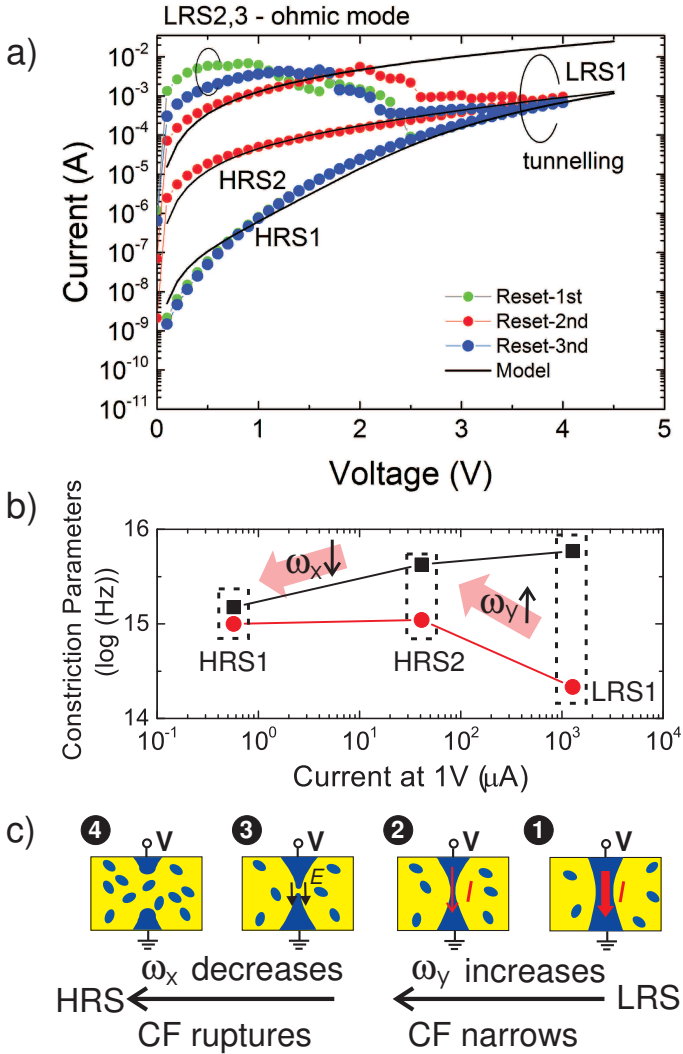}}
\caption{a) Reset I-V curves (symbols) of Au/SLG/ta-C/Pt device with d=20$\mu$m in 3 consecutive switching cycles. The  I-V curves of HRS1, HRS2, and LRS1 are fitted by the QPC model (line). b) Fitting parameters $\omega_x$, $\omega_y$ as a function of current at 1V. V$_0$ is 0.15eV for all cases. c) Schematic of 4 stages when the device is reset from LRS to HRS.}
\label{fig:Fig7}
\end{figure}

We also observe multiple resistive states in the device with SLG, Fig.\ref{fig:Fig7}. We analyze these with the QPC model. Fig.\ref{fig:Fig7}a plots the measured I-V curves over 3 consecutive RESET processes for the same d=20$\mu$m. The SET processes are achieved under the same conditions by sweeping the voltage to Pt from 0 to 6V with I$_{CC}=$3mA. Due to the stochastic nature of the reversible, soft-breakdown, i.e. the CF formation, in the SET process, 3 LRS states can be observed. While LRS2 and LRS3 have a linear I-V relation and are governed by Ohm's law, LRS1 has a non-linear I-V curve and can be fitted by the QPC model. Two HRS states can be seen in Fig.\ref{fig:Fig7}a. Although the device is reset to HRS1 in the 1$^{\text{st}}$ and 3$^{\text{rd}}$ cycle, it is reset to an intermediate HRS2 in the 2$^{\text{nd}}$ cycle. Both HRS1 and HRS2 can be fitted by the QPC model, as shown by the black line in Fig.\ref{fig:Fig7}a. Fig.\ref{fig:Fig7}b plots the fitting parameters used to describe the dimensions of the constriction in LRS1, HRS1 and HRS2. 4 CF representative stages at different resistive states in the RESET process are sketched in Fig.\ref{fig:Fig7}c. LRS1 corresponds to stage 1: a wide CF. In HRS2, $\omega_y$ is increased compared to LRS1, which leads to a gradual narrowing of the CF. This indicates a decrease of the CF cross-section, as in stage 2. Furthermore, compared to the fitting parameters for HRS2, $\omega_x$ at HRS1 is smaller, while $\omega_y$ remains almost constant. This indicates the decrease of the longitudinal constriction until the CF ruptures, stage 4. This reveals the dynamics of the RESET process. The CF starts to become thinner and then ruptures due to the fuse-effect caused by the strong local Joule heat at temperatures of $>1500$K for the SET and between 1500 and 2500K for RESET as calculated by Ref.\cite{Koelmans2016}. Overall, the QPC model can be used to describe the I-V curves and sheds light on the switching mechanism. It indicates that the reduction in leakage current is due to the SLG, paving the way for the development of 4F$^2$ sized cells.
\section{\label{Con} Conclusions}
We investigated the resistive switching properties of ta-C based MIM devices, including forming-free process, polarity-independent operation, ON/OFF ratio and power density. We showed that, by adopting graphene-based electrodes, the ON/OFF ratio is improved by one order of magnitude without increasing the switching power. We attribute the underlying mechanism to a suppressed tunneling current due to the low DOS of graphene near the Dirac point. We also demonstrated the capability of multilevel storage. This paves the way for RRAMs integrated in crossbar arrays without need of an access transistor, and the development of 4F$^2$ cells.
\section{acknowledgements}
We acknowledge funding from EU project CareRAMM, EU Graphene Flagship, ERC Grant Hetero2D, EPSRC Grants EP/509K01711X/1, EP/K017144/1, EP/N010345/1, EP/M507799/ 5101, EP/L016087/1.

\end{document}